\input{aipcheck}

\documentclass[
    ,final            
  ]
  {aipproc}

\layoutstyle{6x9}

\newcommand{\Msun}{M_\odot}

\newcommand{\amin}{a_{\rm crit}}
\newcommand{\SPGP}{SPGP}

\begin{document}

\title{Orbital Evolution of Planets around Intermediate-Mass Giants}

\classification{97.82.Fs, 96.15.Bc, 96.15.Wx}
\keywords      {exoplanet, planetary system--formation and evolution, tide}

\author{M. Kunitomo}{
  address={Department of Earth and Planetary Sciences, Tokyo Institute of Technology, 2-12-1 Ookayama, Meguro-ku, Tokyo 152-8551, Japan}
}
\author{M. Ikoma}{
  address={Department of Earth and Planetary Sciences, Tokyo Institute of Technology, 2-12-1 Ookayama, Meguro-ku, Tokyo 152-8551, Japan}
}
\author{B. Sato}{
  address={Department of Earth and Planetary Sciences, Tokyo Institute of Technology, 2-12-1 Ookayama, Meguro-ku, Tokyo 152-8551, Japan}
}
\author{Y. Katsuta}{
  address={Department of Cosmosciences, Hokkaido University, Kita 10 Nishi 8, Kita-ku, Sapporo 060-0810, Japan}
}
\author{S. Ida}{
  address={Department of Earth and Planetary Sciences, Tokyo Institute of Technology, 2-12-1 Ookayama, Meguro-ku, Tokyo 152-8551, Japan}
}

\begin{abstract}
Around low- and intermediate-mass (1.5-3 $\Msun$) red giants, no planets have been found inside 0.6 AU. 
Such a paucity is not seen in the case of 1$\Msun$ main sequence stars. 
In this study, we examine the possibility that short-period planets were engulfed by their host star evolving off the main sequence. 
To do so, we have simulated the orbital evolution of planets, including the effects of stellar tide and mass loss, 
to determine the critical semimajor axis, $\amin$, beyond which planets survive the RGB expansion of their host star. 
We have found that $\amin$ changes drastically around 2 $\Msun$: 
In the lower-mass range, $\amin$ is more than 1~AU, while $\amin$ is as small as about 0.2~AU in the higher-mass range. 
Comparison with measured semimajor axes of known planets suggests that there is a lack of planets 
that only planet engulfment never accounts for in the higher-mass range. 
Whether the lack is real affects our understanding of planet formation. 
Therefore, increasing the number of planet samples around evolved intermediate-mass stars is quite meaningful to confirm robustness of the lack of planets.
\end{abstract}

\maketitle


\section{INTRODUCTION}

According to recent radial velocity surveys for GK giants, 
it appears that there is a lack of giant planets inside 0.6~AU \citep{Sato+08a, Omiya+09}. 
Such a paucity is not seen around Sun-like stars which often harbor short-period planets such as hot Jupiters.  
One possibility is that such a property is primordial; namely, short-period giant planets ({\SPGP}s) are not present around BA dwarfs originally. 
Another possibility is that {\SPGP}s were removed during the evolution of their host star, which is to be explored in this paper.

A star evolves off the main sequence (MS) towards the red-giant branch (RGB), after exhaustion of hydrogen at its center. 
Then, once the central helium ignites, the star enters a next stable phase called the horizontal branch (HB). 
Most of the giants with detected planets are thought to be on their HB. Before reaching the HB, namely, in the RGB phase, 
the star expands substantially, becomes highly luminous, and loses substantial mass, which should affect orbits of surrounding planets.

\citet{Villaver+Livio09}, following \citet{Sato+08a}, examined what happened around
evolving intermediate-mass stars to understand the observed lack of {\SPGP}s around GK giants. 
They demonstrated that Jovian-mass planets initially orbiting at $< \sim$~1~AU around 2-3$\Msun$ stars 
undergo orbital decay due to stellar tide, ending up being swallowed by their host star. 
However, they made no quantitative comparison with observation.

The purpose of this paper is to examine whether planet engulfment by host stars is responsible for the lack of {\SPGP}s, 
by making a quantitative comparison between the limits of existence of {\SPGP}s predicted theoretically and suggested observationally. 
To this end, we derive a critical semimajor axis beyond which planets survive the host star's RGB phase, and investigate its sensitivity to stellar parameters.

\section{PHYSICAL MODEL}

We simulate the evolution of circular orbit of a planet during its host star's evolution. 
The effects of stellar tide and mass loss on planetary orbit are included. 
We neglect other competing processes such as the frictional and gravitational drag forces by stellar wind
and change in the planet mass due to stellar-wind accretion and evaporation, which were evaluated to be negligible by
\citet{Villaver+Livio09} and \citet{Duncan+Lissauer98}. Thus, we integrate
   \begin{equation}
     \frac{1}{a} \frac{\mathrm{d}a}{\mathrm{d}t} = 
        -6 \frac{k}{T} \frac{M_p}{M_\star} 
	               \left(1+\frac{M_p}{M_\star}\right)
		       \left(\frac{R_\star}{a}\right)^8
	-\frac{\dot{M}_\star}{M_\star},
   \label{eq:orbit}
   \end{equation}
where $a$ is the semimajor axis, $t$ is time, $M_p$ is the planet's mass, 
$M_\star$ and $R_\star$ are the mass and radius of the host star, respectively, 
and $k$ is the apsidal motion constant, and $T$ is the eddy turnover timescale.
The first term on the right-hand side represents the effect of stellar tide \citep{Hut81}. 
Since an RGB star is a slow rotator, we assume no stellar rotation, 
which means that the stellar tide always causes orbital decay of the planet. 
As for the parameters for stellar tide such as $k$ and $T$, 
we have followed \citet{Rasio+96} and \citet{Villaver+Livio09}.

The second term on the right-hand side of equation~(\ref{eq:orbit}) represents
orbital migration due to stellar mass loss ($\dot{M}_\star < 0$). We
use the Reimers' parameterisation for stellar mass loss
\citep{Reimers75}, namely,
    $ \dot{M}_\star = - 4 \times 10^{-13} \eta {L_\star R_\star}/{M_\star}$,
where $\eta$ is the mass-loss parameter; $\dot{M}_\star$ is in $\Msun
\, {\rm yr}^{-1}$ and $M_\star$, $L_\star$, and $R_\star$ are in solar
units.

We simulate stellar evolution directly with the code
MESA~v2258 \citep{Paxton+10} to calculate $M_\star$,
$R_\star$, $\dot{M}_\star$, and parameters relevant to stellar tide
as a function of time $t$. In this calculation in this paper, the effect of convective overshooting is not included (see \citet{Kunitomo+11} for the
results with overshooting).

\section{ORBITAL EVOLUTION AND SURVIVAL LIMIT \label{sec:results}}
\begin{figure}[tbh]
	\includegraphics[width=11.5cm,keepaspectratio]{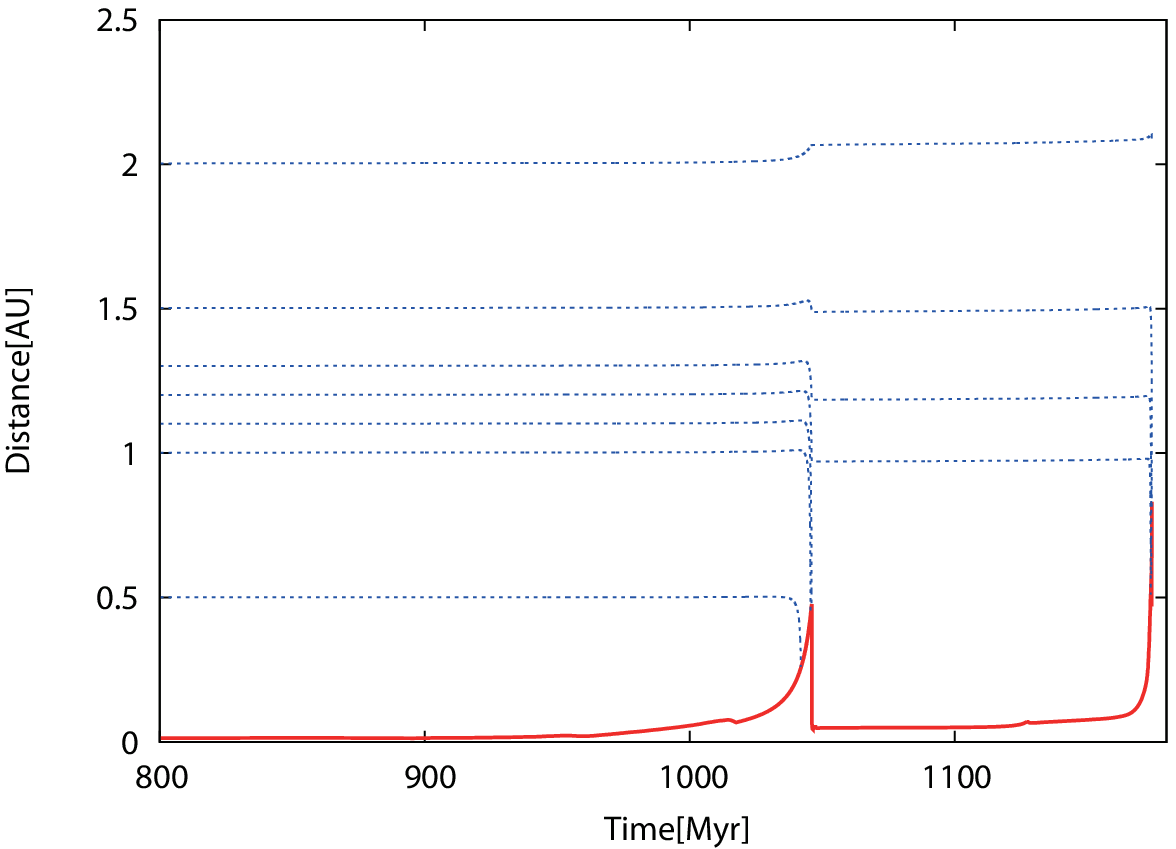} 
\end{figure}
\begin{figure}[tbh]
	\includegraphics[width=11.5cm,keepaspectratio]{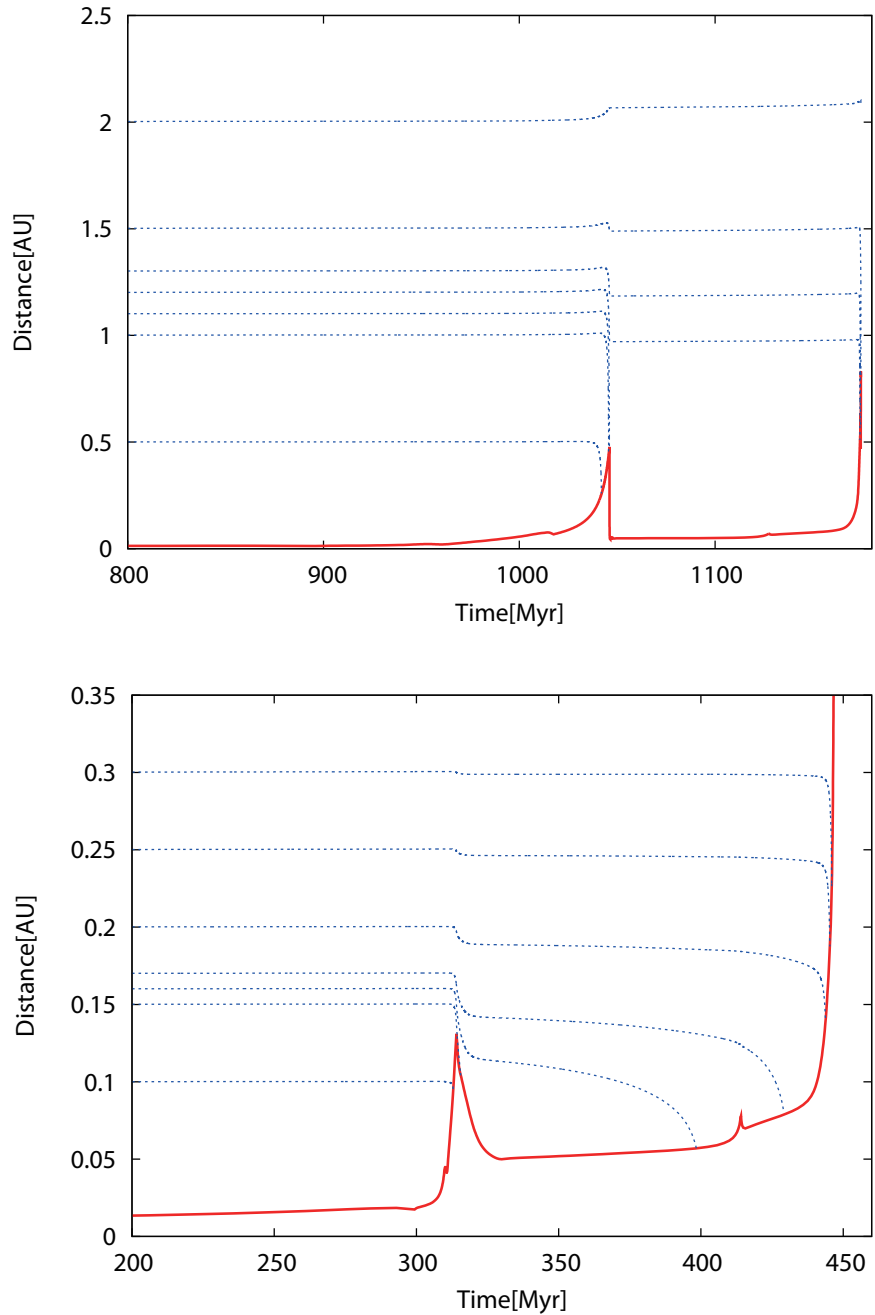}
	\caption{Evolution of semimajor axes of  Jupiter-mass planets (dotted lines) 
		and the radius of the solar-metallicity star (solid line) 
		for $M_\star$ = 2~$\Msun$ (upper panel) and 3~$\Msun$ (bottom panel). \label{fig:orbital evolution}}
\end{figure}

Figure~\ref{fig:orbital evolution} shows the orbital evolution of
planets around stars with masses of $2 \Msun$ (upper panel) and $3
\Msun$ (lower panel); the metallicity is solar ($Z_\star = 0.02$) in
the calculations. The solid lines represent the evolution of the
stellar radius $R_\star$.  The orbit of a 1$M_J$ planet with different
initial semimajor axes (dotted lines) has been integrated from the
host-star's zero-age main sequence until the planet's semimajor axis
falls below the stellar radius ($a < R_\star$) or until the beginning of Asymptotic Giant Branch (AGB) thermal pulse.

Of special interest in this study is whether planets survive the RGB and HB phases of their host star. 
As seen in the upper panel of Fig.~\ref{fig:orbital evolution}, the 2$\Msun$ star expands up to $0.5$~AU at the RGB-tip 
and consequently swallows planets whose initial semimajor axes, $a_{i}$, are smaller than 1.1~AU.  
The planet that starts out at 1.2~AU is pulled by the host star, but it barely survives the RGB/HB phases. 
We define a critical initial semimajor axis, $\amin$, below which planets end up being engulfed by their host star at some point on the RGB or HB. 
In the present case, $\amin = 1.2~{\rm AU}$. (Hereafter, we sometimes call the critical initial semimajor axis the survival limit.)
As for the 3$\Msun$ star (the lower panel of Fig.~\ref{fig:orbital
evolution}), the RGB-tip radius is as small as 0.13~AU.  As a
consequence, even a planet starting out at 0.20~AU can avoid
engulfment during the RGB and HB phases.
As seen in these panels, the survival limit is sensitive to the mass of host stars. 
While $\amin = 1.2$~AU for the 2$\Msun$ star, $\amin = 0.20$~AU for the 3$\Msun$ star.

\begin{figure}[t]
    \includegraphics[width=12cm,keepaspectratio]{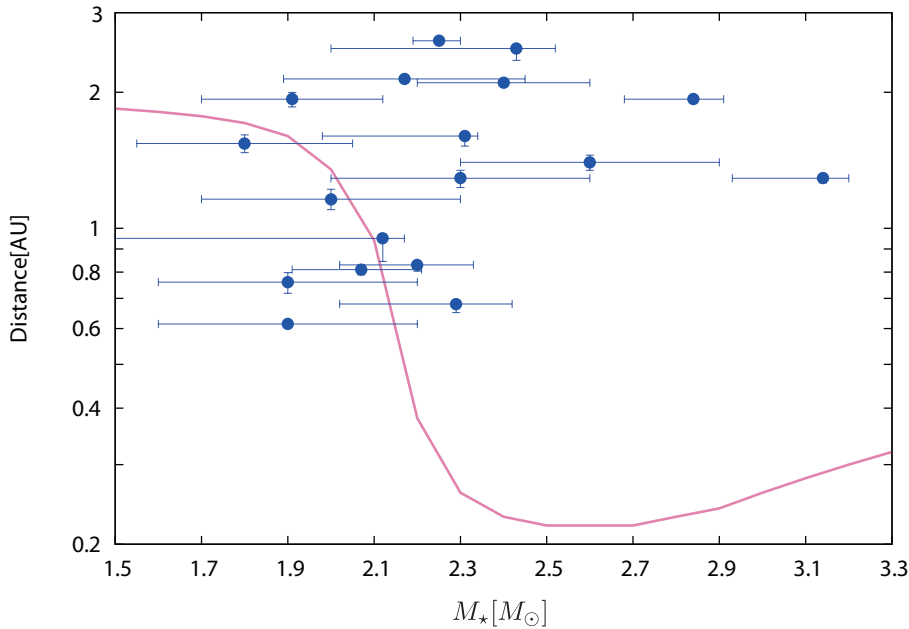}
	\caption{Comparison between the critical semimajor axis $\amin$ (solid line) 
	and measured semimajor axes of detected exoplanets (symbols with error bars). 
	The theoretical curve is for stellar metallicity $Z = 0.01$ and planetary mass $M_p = 6 M_J$. In this calculation, the effect of overshooting is not included. \label{fig:observation}}
\end{figure}

\section{COMPARISON WITH OBSERVATION \label{sec:discussion}}
We compare measured semimajor axes of known planets around GK giants with the critical semimajor axis that we have derived in Fig.~\ref{fig:observation}.
Those measured stellar masses and planetary semimajor axes are taken from the literature 
\citep[e.g.,][]{Takeda+08}.
Two facts can be found in this figure: First, the observational errors
being taken into account, it would be fair to say that all the known
planets are outside the survival limit. Therefore, it is not denied
that {\SPGP}s ($a < \amin$) were swallowed by their host star on the
RGB. Second, around giants of $M_\star > 2.4 \Msun$, planets exist far
from $\amin$. In other words, planet engulfment by host stars seems
not to be the main reason for the lack of {\SPGP}s.

Another possibility is that the lack of {\SPGP}s is primordial---there
may be any stellar-mass-dependent processes that hinder formation of
{\SPGP}s around 2-3$\Msun$ stars.  For example, \citet{Burkert+Ida07}
proposed that a decreasing-with-stellar-mass lifetime of
protoplanetary disks could become comparable with or shorter than the
timescale of the type-II migration of giant planets. They then
demonstrated that the observed period valley between the hot-Jupiter
and cool-Jupiter classes is more pronounced for planets orbiting F
stars ($1.2$-$1.5\Msun$) than GK stars ($0.8$-$1.2\Msun$). Applying
the idea to high-mass stars ($1.5$-$3.0\Msun$), \citet{Currie09}
demonstrated that {\SPGP}s are rarely formed around high-mass stars.

Obviously we need more samples of giant planets and smaller planets orbiting high-mass stars to confirm conclusions we have derived in this paper and to 
understand stellar-mass-dependence of planet formation, which is also helpful in understanding the origins and diversity of planetary systems around solar-type stars. Therefore, further surveys for planets around giants are highly encouraged.


\begin{theacknowledgments}
We would like to express our gratitude to the following persons: B. Paxton and A. Dotter kindly helped us install and use the stellar-evolution code MESA and modified it upon our request. M. Fujimoto and T. Suda gave useful comments about stellar evolution. We had fruitful discussion on
this study with Y. Hori and T. Nakamoto. This work is supported partly by Japan Society for the Promotion of Science (JSPS).
\end{theacknowledgments}


\begin{thebibliography}{}
  \bibitem[Burkert \& Ida(2007)]{Burkert+Ida07} Burkert, A., \& Ida, S. 2007, ApJ, 660, 845 
  \bibitem[Currie(2009)]{Currie09} Currie, T. 2009, ApJ, 694, L171 
  \bibitem[de medeiros et al.(2009)]{de medeiros+09} de Medeiros, J.R., et al.\ 2009, A\&A, 504, 617
  \bibitem[Duncan \& Lissauer(1998)]{Duncan+Lissauer98} Duncan, M.~J., \& Lissauer, J.~J.\ 1998, Icarus, 134, 303 
  \bibitem[Dollinger et al.(2009)]{Dollinger+09} D\"ollinger, M.P., Hatzes, A.P., Pasquini, L., Guenther, E.W., \& Hartmann, M.\ 2009, A\&A, 505, 1311
  \bibitem[Hatzes et al.(2006)]{Hatzes+06} Hatzes, A.P., et al.\ 2006, A\&A, 457, 335
  \bibitem[Hut(1981)]{Hut81} Hut, P.\ 1981, A\&A, 99, 126
  \bibitem[Liu et al.(2009)]{Liu+09} Liu, Y.-J., Sato, B., Zhao, G., \& Ando, H.\ 2009, RAA, 9, L1
  \bibitem[Lovis \& Mayor(2007)]{Lovis+Mayor07} Lovis, C., \& Mayor, M.\ 2007, A\&A, 472, 657
  \bibitem[Kunitomo et al.(2011)]{Kunitomo+11} Kunitomo, M., Ikoma, M., Sato, B., Katsuta, Y., \& Ida, S.\ 2011, ApJ, submitted
  \bibitem[Niedzielski et al.(2007)]{Niedzielski+07} Niedzielski, A., et al.\ 2007, ApJ, 669, 1354
  \bibitem[Niedzielski et al.(2009a)]{Niedzielski+09a} Niedzielski, A., Gozdziewski, K., Wolszczan, A., Konacki, M., Nowak, G., \& Zielinski, P.\ 2009, ApJ, 693, 276
  \bibitem[Omiya et al.(2009)]{Omiya+09} Omiya, M., et al.\ 2009, PASJ, 61, 825 
  \bibitem[Paxton et al.(2010)]{Paxton+10} Paxton, B., Bildsten, L., Dotter, A., Herwig, F., Lesaffre, P., \& Timmes, F.\ 2010, arXiv:1009.1622
  \bibitem[Rasio et al.(1996)]{Rasio+96} Rasio, F.~A., Tout, C.~A., Lubow, S.~H., \& Livio, M.\ 1996, ApJ, 470, 1187 
  \bibitem[Reimers(1975)]{Reimers75} Reimers, D. 1975, in Problems in Stellar Atmospheres and Envelopes, ed. B. Bascheck,W.~H. Kegel, \& G. Traving (New York: Springer), 229
  \bibitem[Sato et al.(2007)]{Sato+07} Sato, B., et al.\ 2007, ApJ, 661, 527
  \bibitem[Sato et al.(2008a)]{Sato+08a} Sato, B., et al.\ 2008, PASJ, 60, 539
  \bibitem[Sato et al.(2008b)]{Sato+08b} Sato, B., et al.\ 2008b, PASJ, 60, 1317
  \bibitem[Sato et al.(2010)]{Sato+10} Sato, B., et al.\ 2010, PASJ, 62, 1063
  \bibitem[Setiawan et al.(2005)]{Setiawan+05} Setiawan, J., et al.\ 2005, A\&A, 437, L31
  \bibitem[Takeda et al.(2008)]{Takeda+08} Takeda, Y., Sato, B., \& Murata, D.\ 2008, PASJ, 60, 781
  \bibitem[Villaver \& Livio(2009)]{Villaver+Livio09} Villaver, E., \& Livio, M.\ 2009, ApJ, 705, L81 
\end{thebibliography}
\end{document}